# Terahertz Plasmonic Transport in Topological Valley Metal-slabs

Xiang Zhou[a], Hui-Chang Li[a], Yun Shen*[a,b]

[a]Department of Physics, Nanchang university, Nanchang 330031, China
[b]Institute of Space Science and Technology, Nanchang university, Nanchang 330031, China

**Abstract**

Topological photonic devices have attracted great attentions in terahertz (THz) and optical regimes due to their robust protected transport properties. However, it remains challenging in miniaturization of the devices to get superior performance for photonic integrated circuits in optical networks. In this paper, Kagome photonic insulators constructed with ultrathin metal-slab on Polyimide substrate are proposed for THz waveguiding. Theoretical analysis and numerical simulation demonstrate that $C_{3v}$ symmetry can be broken by global rotation $\theta$ of the air holes in metallic Kagome lattice, providing topological phase transitions. The propagation of THz waves through Z-shaped domain walls with multiple sharp corners verifies the robustness of plasmonic transport. The positive/negative refraction of topological valley edge state from Zigzag interface into background space is illustrated. These results present a novel approach to manipulate THz waves and facilitate development of photonic integrated circuits with high compactness and robustness.

## 1. Introduction

Terahertz (THz) wave has received rapidly growing attentions due to great potential in offering a broader available bandwidth for higher data transfer rates [1-6]. Specifically, THz waveguides are required for realization of high-speed sixth-generation (6G) networks, interconnects of intrachip/interchip communication, THz integrated circuits, and so on [7-11]. However, conventional approaches for THz waveguiding are generally sensitive to defects, disorder and bending at sharp corners [12-16]. To overcome such obstacles, THz photonic topological insulator is proposed and experimentally investigated in different structures [17-20]. In which, robust energy transport and strong suppression of backscattering are provided due to topological phase transitions of broken symmetry. More recently, tunable THz topological edge and corner states in spoof surface plasmon crystals are also studied [21-23]. It shows that in two-dimensional metallic cylinders structures, topological robustness can be present. Nonetheless, challenging remains in miniaturization of the devices to get better performance for photonic integrated circuits in optical networks.

In this paper, Kagome photonic insulators constructed with 100 nm thickness metal-slab on Polyimide (PI) substrate are proposed for THz waveguiding. By rotating the global $\theta$ of air holes in metallic Kagome lattice, $C_{3v}$ symmetry can be broken and provide topological phase transitions. THz wave transmissions are verified with straight and Z-shaped topological waveguides, which quantitatively demonstrate the strong backscattering suppression effects. Furthermore, positive/negative refraction of topological valley edge state from the Zigzag interface into background space is investigated. Such THz topological plasmonic insulators may serve as efficient waveguide for high-speed, on-chip communication devices, while the ultrathin thickness and planar feature would facilitate integration of the photonic circuits.

## 2. Design of valley photonic insulator and band structures

The metal-slab Kagome photonic insulator is shown in Fig. 1(a). In which, the unit cell of Kagome lattice is composed of three identical rectangular air holes with width $W = 95$ μm, length $L = 50$ μm, height $d = 0.1$ μm and lattice constant $a = 200$ μm, respectively. The substrate is PI with permittivity of $\varepsilon \approx 3.5$ and height $h = 50$ μm.

The global rotational angle of the air holes in Kagome lattice unit cell is denoted by $\theta$. As $\theta = 0°, 30°$ and $60°$, the photonic bands of Kagome photonic crystals (PCs) are illustrated in Fig. 1(b). It shows that the bandgap closes at valley $K(K')$ for $\theta = 30°$ due to $C_{3v}$ symmetry, and open for $\theta = 0°$ and $60°$ implying the broken-symmetry and phase transitions [24-26]. The $E_z$ field variations in the first and second bands at $K$ point are severally depicted in Fig. 1(c) for $\theta = 0°$ and $60°$. In which, the either left or right circularly polarized energy flux of the states imply the inversion.

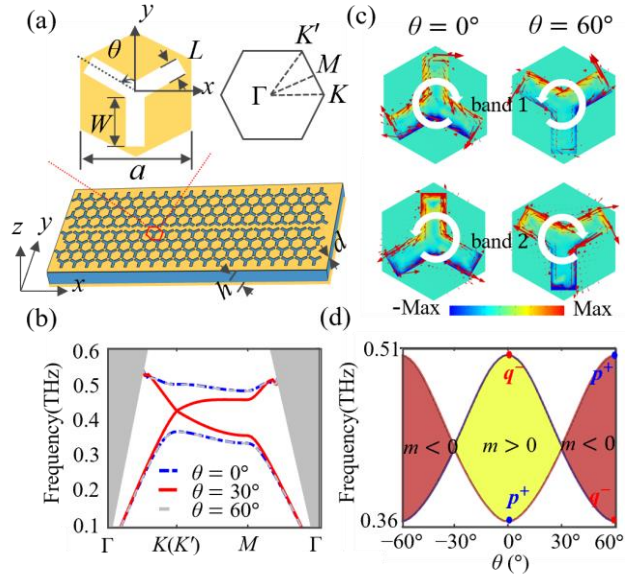

**Fig. 1** Plasmonic Kagome lattice PCs and phase transitions. (a) Schematic of plasmonic Kagome lattice PCs. (b) Photonic bands for $\theta = 0°, 30°$ and $60°$. (c) Field distributions of the two eigenstates at $K$ point for $\theta = 0°, 60°$. (d) Dependence of topological phase transitions on $\theta$ at $K$ point; the color regions represent bandgaps with effective mass $m >$

0 and $m < 0$, respectively.

Theoretically, valley Hall phase transition in the plasmonic crystals can be described by an effective Hamiltonian through the **k·p** perturbation method [27, 28]. Near valley $K$, it can be written as:

$$H_K(\delta \bm{k}) = v_D \delta k_x \sigma_x + v_D \delta k_y \sigma_y + m v_D^2 \sigma_z \tag{1}$$

where $v_D$, $\delta \bm{k}$, and $\sigma$ are group velocity at degenerate Dirac cone, momentum deviation from the $K$ point, and Pauli matrix, respectively. $m = \pi(q-p)/2v_D^2$ with $q$ and $p$ severally indicating the two frequencies at boundaries of bandgap after Dirac point lifted. Here, we note that the band inversion and topological phase transition can also be derived from the flips of $p$ and $q$ states shown in Fig. 1(d). In which, dependence of topological phase transition on $\theta$ at $K$ point is present, and the color regions represent bandgaps with effective mass $m > 0$ and $m < 0$, respectively. Furtherly, the Berry curvature can be derived from effective Hamiltonian as $\Omega_K(\delta \bm{k}) = (mv_D)/2(\delta \bm{k}^2 + m^2 v_D^2)^{3/2}$. The valley Chern number can be subsequently calculated with:

$$C_K = \frac{1}{2\pi} \int \Omega_K(\delta \bm{k}) d\bm{k}^2 = \pm \frac{1}{2} \text{sgn}(m) \tag{2}$$

The equation (2) shows $C_K$ is only determined by the sign of $m$.

**3. Topological valley edge states**

To investigate topological states, the distributions of Berry curvature of the first band for our proposed plasmonic crystals with $\theta = 0°$ and $60°$ are severally numerically calculated by COMSOL Multiphysics and shown in Fig. 2(a). As we know that valley edge states exist at $|\Delta C^{K/K'}| = 1$, the possible edge states at domain wall separated by the $\theta = 0°$ and $\theta = 60°$ PCs are illustrated in Figs. 2(b)(i) and

2(b)(ii). Correspondingly, the $|E_z|$ field distributions and band diagrams for $0°/60°$ and $60°/0°$ are demonstrated in Figs. 2(c)(i) and 2(c)(ii).

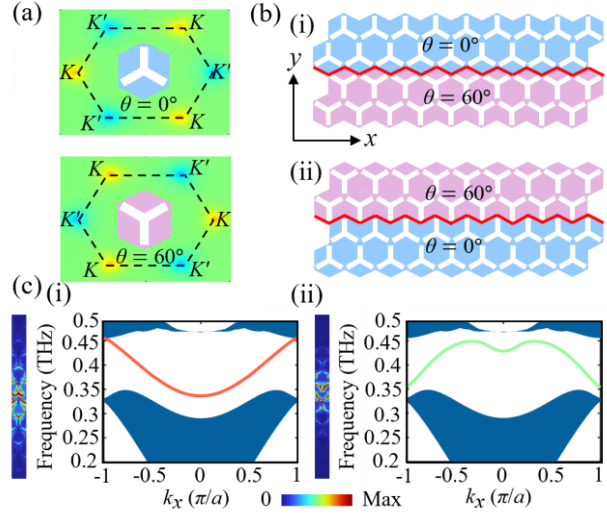

**Fig. 2** Properties of edge states. (a) Distribution of Berry curvatures for plasmonic crystals with $\theta = 0°$ and $\theta = 60°$, respectively. (b) Possible edge states at domain wall separated by PCs of $\theta = 0°$ and $60°$ (c) $|E_z|$ field distributions and band diagrams for $0°/60°$ and $60°/0°$.

### 3.1. Robust topological transport

To demonstrate the robust transport properties of topological edge states [29, 30], both straight-line and Z-shaped topological waveguides are explored by CST Studio Suite. Specifically, the $|E_z|$ field distributions of straight-line, Z-shaped with two turns, and Z-shaped with ten turns for 0.4 THz incidences are illustrated in Fig. 3(a), (b) and (c), respectively. Meanwhile, the corresponding transmissions of the three waveguides, as well as the bulk PC structure, are depicted in Fig. 3(d). In which, a distinct dip (grey curve) from 0.38 to 0.47 THz appears, implying the bulk bandgap. Nonetheless, the transmissions of straight, and Z domain walls with two and ten turns are similar in range of 0.38–0.44 THz, indicating that length and sharp corners of the domain wall have less influences on the transmission. That is, topological edge states, which are immune to sharp corner scattering, are provided.

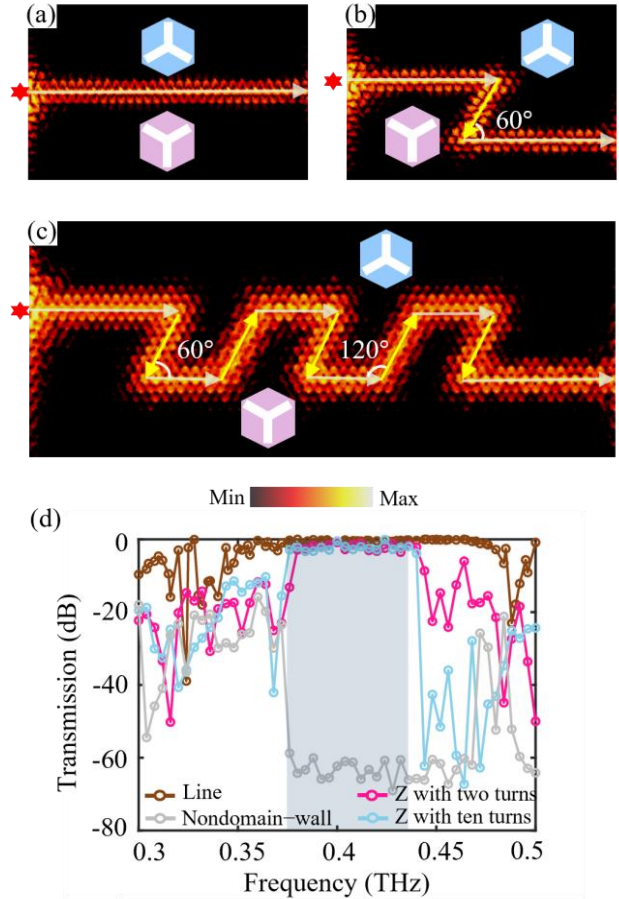

**Fig. 3** Transmission. (a)-(c) $|E_z|$ field distributions of (a) straight-line, (b) Z-shaped with two turns, and (c) Z-shaped with ten turns for 0.4 THz incidences. (d) Calculated transmissions of straight lines, Z-shaped domain walls, and bulk.

To furtherly study the robustness of the topological edge states, disorder and air defects are introduced into structures [31, 32]. The configurations and $|E_z|$ field distributions for 0.4 THz incidences are illustrated in Fig. 4(a)-(b), respectively. The corresponding transmissions of the configurations, as well as bulk PC structure, are shown in Fig. 4(c), respectively. In which, the transmissions for defects in straight-line and Z are similar in range of 0.38–0.44 THz, indicating the robustness.

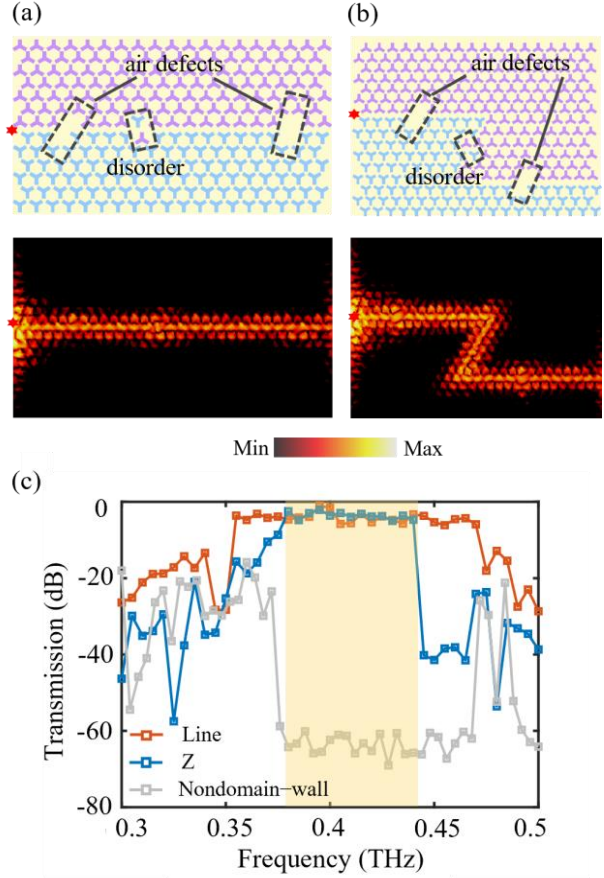

**Fig. 4** Transmission for defects. (a)-(b) Configurations with defects and $|E_z|$ field distributions of (a) straight-line, (b) Z-shaped for 0.4 THz incidences. (c) Calculated transmissions of straight lines, Z-shaped, and bulk.

### 3.2. Valley topological refractions

For potential applications in photonic integrated circuits, the valley topological refractions of the plasmonic waveguides are also explored here [33-35]. As shown in Fig. 5(a) and (b), the incident $k$ can be matched with the background equifrequency curve to determine the direction of the outgoing beam. Specifically, by virtue of the phase-matching condition $k \cdot e_{\text{term}} = K \cdot e_{\text{term}}$, the theoretical refracted angles can be calculated by $|k|\cos(120° + \theta_{r1}) = |K|\cos 60°$ and $|k|\cos(60° - \theta_{r2}) = |K'|\cos 60°$, respectively. For 0.44 THz and 0.42 THz, $\theta_{r1} = -67.3°$ and $\theta_{r2} = 16.4°$ can be obtained, and the $|E_z|$ field are severally illustrated in (a) and (b), exhibiting negative and positive refractions.

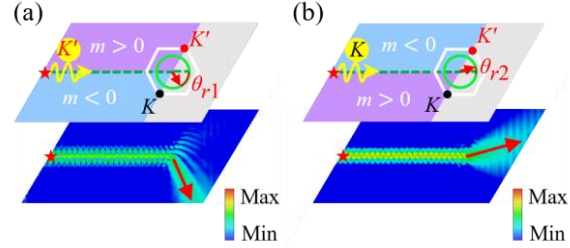

**Fig. 5** Valley topological refractions. The *k*-space analyses on out-coupling of (a) *K'* and (b) *K* valley projected edge states along the negative- and positive-type Zigzag interface, respectively.

## 4. Conclusion

In summary, Kagome photonic insulators constructed with ultrathin metal-slab are proposed for THz waveguiding. Plasmonic transports for topological edge states of straight-line, Z-shaped with two turns, Z-shaped with ten turns, and configurations with defects are studied. The results quantitatively demonstrate the strong backscattering suppression effects. In addition, the valley topological refractions are explored for potential applications in photonic integrated circuits. Such THz topological plasmonic insulators present a novel approach to manipulate THz waves and facilitate development of integrated devices with high compactness, and robustness, serving for 6G networks, interconnects of intrachip/interchip communication, THz integrated circuits, etc.

**Acknowledgements**

This research was supported by the National Natural Science Foundation of China (Grant numbers 61927813, 61865009).